\documentclass[prl, reprint, nofootinbib, 10pt, showkeys, showkeys]{revtex4-2} 
\usepackage{etex}

\usepackage{amsmath, amssymb, amsfonts, braket, commath, bm, empheq, array, lipsum} 

\usepackage[]{graphicx}
\graphicspath{{images/}}
\DeclareGraphicsExtensions{.eps,.jpg,.png,.pdf}
\usepackage{hyperref, url}
\hypersetup{
	colorlinks=true,
	linkcolor=red,
	filecolor=blue,
	urlcolor=magenta,
}

\usepackage{dblfloatfix}

\newcommand{\Ai}[1]{\ensuremath{\mathrm{Ai}\del{#1}}}
\newcommand{\bte}{\ensuremath{\beta\textrm{ ensemble}}}

\newcommand{\hQHO}{ \ensuremath{\hat{H}_Q}}						
\newcommand{\ipr}{\ensuremath{\textrm{I}}}

\newcommand{\nrm}{ \ensuremath{\mathcal{Z}} }
\newcommand{\nmax}{ \ensuremath{n_0} }							

\begin{document}
\title{Emergent multifractality in power-law decaying eigenstates}

\author{Adway Kumar Das}\email{akd19rs062@iiserkol.ac.in}
\affiliation{
	Indian Institute of Science Education and Research Kolkata, Mohanpur, 741246 India
}
\author{Anandamohan Ghosh}\email{anandamohan@iiserkol.ac.in}
\affiliation{
	Indian Institute of Science Education and Research Kolkata, Mohanpur, 741246 India
}
\author{Ivan M. Khaymovich}\email{ivan.khaymovich@gmail.com}
\affiliation{Nordita, Stockholm University and KTH Royal Institute of Technology Hannes Alfv\'ens v\"ag 12, Sx-106 91 Stockholm, Sweden}
\affiliation{Institute for Physics of Microstructures, Russian Academy of Sciences, 603950 Nizhny Novgorod, GSP-105, Russia}

\date{\today}
\begin{abstract}
Eigenstate multifractality is of significant interest with potential applications in various fields of quantum physics.
Most of the previous studies concentrated on fine-tuned quantum models to realize multifractality which is generally believed to be a critical phenomenon and fragile to random perturbations.
In this work, we propose a set of generic principles based on the power-law decay of the eigenstates which allow us to distinguish a fractal phase from a genuine multifractal phase.
We demonstrate the above principles in a 1d tight-binding model with inhomogeneous nearest-neighbor hopping that can be mapped to the standard quantum harmonic oscillator via energy-coordinate duality.
We analytically calculate the fractal dimensions and the spectrum of fractal dimensions which are in agreement with numerical simulations.
\end{abstract}
\pacs{05.45.Mt}	
\pacs{02.10.Yn} 
\pacs{89.75.Da} 
\keywords{Multifractality, power-law decay, quantum harmonic oscillator, spectrum of fractal dimensions}
\maketitle
Quantum ergodicity and its breakdown via emergent local integrals of motion are important for understanding the phenomenon of many-body localization (MBL)~\cite{Basko2006, gornyi2005interacting, Pal2010, Alet2018CRP, Abanin2019RMP, TorresHerrera2013, TorresHerrera2015}. Despite being localized in real space, MBL eigenstates simultaneously exhibit multifractality in the Hilbert space of dimension $N$, i.e.~occupy an extensive volume $\propto N^D$ ($0<D<1$) but a measure zero fraction of the entire Hilbert space~\cite{Luitz2015, Mace_Laflorencie2019_XXZ,Tikhonov2018MBL_long-range,QIsing_2021, Solorzano2021, BastarracheaMagnani2024}. Although widely observed in many-body systems, multifractality is rare in random-matrix models. It appears as a critical phenomenon at the Anderson localization transition~\cite{Anderson1958, Evers2008, PLRBM} and in quasiperiodic systems ~\cite{Cai2013AA+p-wave,DeGottardi2013AA+p-wave,Wang2016AA+p-wave} which are fragile to any uncorrelated disorder. Robust non-ergodic extended phase in random-matrix models typically requires long-range coupling, e.g.~Rosenzweig-Porter model~\cite{RP,Kravtsov_NJP2015,Biroli_RP,Ossipov_EPL2016_H+V,vonSoosten2017non,Monthus2017multifractality,BogomolnyRP2018,LN-RP_RRG,LN-RP_WE,Biroli2021levy,LN-RP_K(w),Buijsman2022circular,Venturelli2023replica,DeTomasi2022nH-RP,sarkar2023fract-RP, Das2019, Das2022, Das2023, VallejoFabila2024}.
However, in all such models, non-ergodic phase hosts not multifractal, but fractal states~\cite{Kutlin2024}, characterized by the fractal dimension $D_q = D$ for all integers $q>0$. In general, $D_q$ dictates the scaling of the $q$th moments of wave-function intensity
\begin{gather}\label{eq:IPR_q}
\ipr_q = \sum_{\vec{n}} |\Psi_E(\vec{n})|^{2q} \sim N^{(1-q)D_q}
\end{gather}
at energy $E$ and the sum goes over the points $\vec{n}$ of a $d$-dimensional lattice of linear size $L$ such that $N=L^d$.

In contrast to the fractal states, genuine multifractality is given by a non-trivial set of fractal dimensions $D_q>D_{q+1}$, $q\geq 1$, leading to many spatial and energy scales~\cite{Evers2008}.
These multiple energy scales give multifractality potential applications in enhancing superconductivity~\cite{Feigelman2007SC-enhance, Feigelman2010AoP, Burmistrov2012, Petrovic2016disorder, sacepe2020quantum}, quantum algorithm speed-ups~\cite{smelyanskiy2018non, kechedzhi2018efficient} and black hole physics~\cite{micklitz2019non, Kamenev-talk-1, Kamenev-talk-2}. Other than many-body systems, a robust multifractal phase exists only in the weighted adjacency matrices of random Erd\"os-R\'enyi graphs in a finite energy range~\cite{Cugliandolo2024}. In this Letter, we discuss the main principles to realize genuine multifractal extended states in random matrices.

\begin{figure}[t]
	\centering
	\includegraphics[width=\columnwidth]{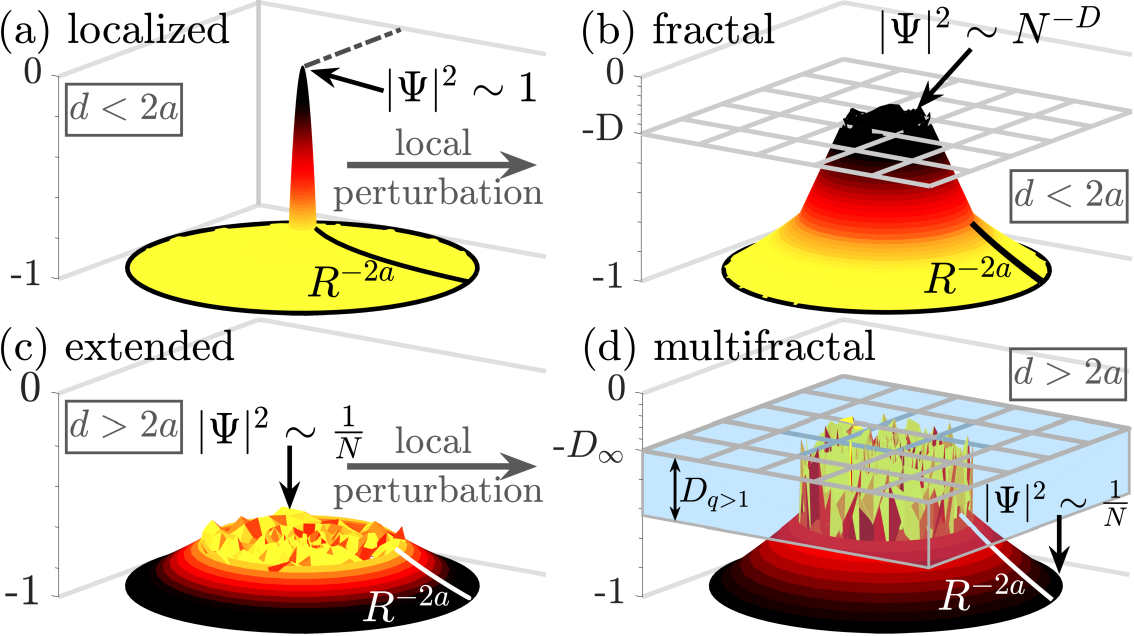}
	\caption{{\bfseries Spatial structures of a $d$-dimensional power-law decaying wave function:} (a)~localized (b)~fractal, (c)~extended, (d)~multifractal.
		For (a-b), $|\Psi|^2 \sim \frac{1}{R^{2a>d}}$, for (c-d), $|\Psi|^2 \sim \frac{1}{R^{2a<d}}$. Dark colors signify the spatial regions mainly contributing to the normalization, the vertical axis shows $\ln|\Psi_E(\vec{n})|^2/\ln N$ with the maximum at $-D_\infty \stackrel{\text{def}}{=}\max_{\vec{n}}\ln|\Psi_E(\vec{n})|^2/\ln N$.
	}
	\label{fig_Cartoon_states}
\end{figure}

Let us consider a generic $d$-dimensional system having eigenstates with a power-law decay
\begin{gather}\label{eq:psi_PL-decay}
	|\Psi_E(\vec{n}_0 + \vec{R})|\sim R^{-a},
\end{gather}
where the maximum intensity occurs at $\vec{n}_0$. For $d<2a$, the normalization condition, $\ipr_1 = 1$ is dominated by a few lattice sites close to the location of the maximum intensity, $|\vec{n}-\vec{n}_0|\lesssim \mathcal{O}(1)$. Consequently, $D_q = 0$ for all integers $q\geq 1$ and the eigenstate is localized, see Fig.~\ref{fig_Cartoon_states}(a). The non-zero fractal dimensions appear only for $q<\frac{d}{2a}<1$, which are determined by the wave-function tails, without affecting localization properties.

Even local perturbations of the localized states, keeping the power-law decaying tail 
at $|\vec{n}-\vec{n}_0|\gg \mathcal{O}(1)$ with $d<2a$, can at most produce fractal states~\cite{Tang2022nonergodic,Motamarri2022RDM}, as illustrated in Fig.~\ref{fig_Cartoon_states}(b).
Since, the maximum eigenstate intensity scales as $|\Psi_E(\vec{n}_0)|^2 \sim N^{-D_\infty}$ where $D_\infty\stackrel{\text{def}}{=}  D_{q\to\infty}$~\cite{Lakshminarayan2008},
for uncorrelated random local perturbation close to $\vec{n}_0$, the wave-function normalization condition implies that $\mathcal{O}(N^{D_\infty})$ number of eigenstate intensities scale similar to $|\Psi_E(\vec{n}_0)|^2$. Consequently, we get fractal states with $D_{q\geq1} = D_\infty>0$ as opposed to power-law localization in Fig.~\ref{fig_Cartoon_states}(a).

In case of power-law decaying localized or fractal states with $d<2a$, the normalization condition $\ipr_1=1$ is dominated by components around the maximum intensity. Contrarily for a slower decay, $d>2a$, the normalization condition is dominated by the tail of the wave-function, $|\vec{n}-\vec{n}_0|\sim \mathcal{O}(L)$, which contains $\mathcal{O}(N)$ typical (most probable) intensities, $|\Psi_{E,typ}|^2 \sim N^{-1}$. In this case, the maximum intensity $|\Psi_E(\vec{n}_0)|^2 \sim N^{-D_\infty}$ barely contributes to the normalization, thus, $D_\infty$ can take any value in $[0, 1]$. In particular, $D_\infty = 1$ corresponds to ergodicity, as shown in Fig.~\ref{fig_Cartoon_states}(c).

For a general value of $D_\infty$, with a power-lay decaying wave function $|\Psi_E(\vec{n}_0 + \vec{R})|\simeq L^{a-d} R^{-a}$ and $q_*\stackrel{\text{def}}{=} \frac{d}{2a}>1$, the state shows genuine multifractal properties, Fig.~\ref{fig_Cartoon_states}(d):
\begin{align}\label{eq:IPR_q_multifractal}
	\begin{split}
		\ipr_q &\simeq \sum_{\vec{R}} \left(\frac{L^{a-d}}{R^a}\right)^{2q} \sim N^{1-q}+\frac{N^{1-q D_\infty}}{N^{q_*(1-D_\infty)}}\\
		\Leftrightarrow D_q &= \begin{cases}
			1 & q\leq q_* = \frac{d}{2a}\\
			1 - (1-D_\infty)\frac{q-q_*}{q-1} & q>q_*>1
		\end{cases}.
	\end{split}
\end{align}
Such states are quite similar to those showing weak multifractality as recently found in random Erd\"os-R\'enyi graphs~\cite{Cugliandolo2024}. Thus, we identify the principle leading to genuine multifractality (fractality) for $d>2a$ ($d<2a$) as a manifestation of uncorrelated random local
perturbations of power-law decaying states, which is one of the main results of this Letter.

The simplest recipe to produce power-law decaying eigenstates in a tight-binding model with on-site disorder is to add 
power-law decaying hopping, e.g.~dipolar Anderson~\cite{Levitov1989,Levitov1990,PLRBM,Mirlin2000RG_PLRBM,Burin1989,Malyshev2000,Malyshev2004,Malyshev2005,Deng2018Duality,Deng2022AnisBM} and random-matrix models~\cite{Cantin2018,Bhatt2018,Nosov2019correlation,Nosov2019mixtures,Kutlin2020_PLE-RG,Kutlin2021emergent}.
But, spatial renormalization group~\cite{Levitov1989,Levitov1990,Mirlin2000RG_PLRBM} dictates that such perturbation breaks down at $a\leq d$ in long-range systems, yielding weak ergodicity for $\frac{d}{2}<a<d$~\cite{BogomolnyPLRBM2018} and ergodicity at $a<\frac{d}{2}$~\cite{vonSoosten2017phase}.
Only in the non-Hermitian setting, power-law decaying eigenstates can survive in the interval $\frac{d}{2}\leq a\leq d$~\cite{DeTomasi2023nH-PLRBM}.
Therefore, long-range systems are unable to exhibit genuine multifractality, which requires $d>2a$.
Thus, we will focus on short-range $1$d models with nearest-neighbor hopping having enough inhomogeneity to avoid exponential localization and will demonstrate that these are sufficient conditions to get genuine multifractality in power-law decaying eigenstates. This is the second important result in this Letter.

Inhomogeneous nearest-neighbor hopping produces many interesting physics e.g.~Anderson localization~\cite{Herbert1971, Thouless1972, TorresHerrera2019}, topological insulators~\cite{Su1979, Chaunsali2021}, perfect information transfer in pre-engineered quantum wires with parabolic hopping~\cite{Albanese2004, Christandl2004, Shi2005, Nikolopoulos2004}, experimentally realizable in photonic qubits~\cite{Chapman2016}. Moreover, Lanczos reduction of many-body Hamiltonians in the Krylov basis~\cite{lanczos1950iteration,arnoldi1951principle,viswanath1994recursion} yields inhomogeneous tridiagonal matrices, where the growth of the correlated hopping elements dictates the transition from integrability to ergodicity~\cite{Parker2019}.

The analogous uncorrelated case is represented by the well-known \bte~\cite{Dumitriu2002}, which is the Krylov-basis representation~\cite{balasubramanian2022tale} of the Wigner-Dyson ensembles for Dyson index $\beta=1$, $2$, and $4$~\cite{Mehta1}.
The \bte\ extends the Dyson index to any real number and
shows many interesting properties: fractal eigenstates at the spectral edge~\cite{Das2024a} and in the bulk~\cite{Das2022a} coexisting with localized states without forming a mobility edge~\cite{Das2023a}, or unusual long-range energy correlation~\cite{Relano2008, Das2022a} at $\beta\ll 1$. At the same time, power-law decay of the characteristic polynomials with non-trivial exponent $a=\tfrac14+\tfrac1{2\beta}$ at large $\beta$ values~\cite{Breuer2007,breuer2010spectral,lambert2023strong_hyperbolic,lambert2021strong_Airy}
may lead to the unusual eigenstate profile.
Thus, \bte\ should be an ideal candidate to produce genuine multifractality in power-law decaying eigenstates, at least for $\beta>1$. 

\begin{figure}[t]
	\centering
	\includegraphics[width=\columnwidth]{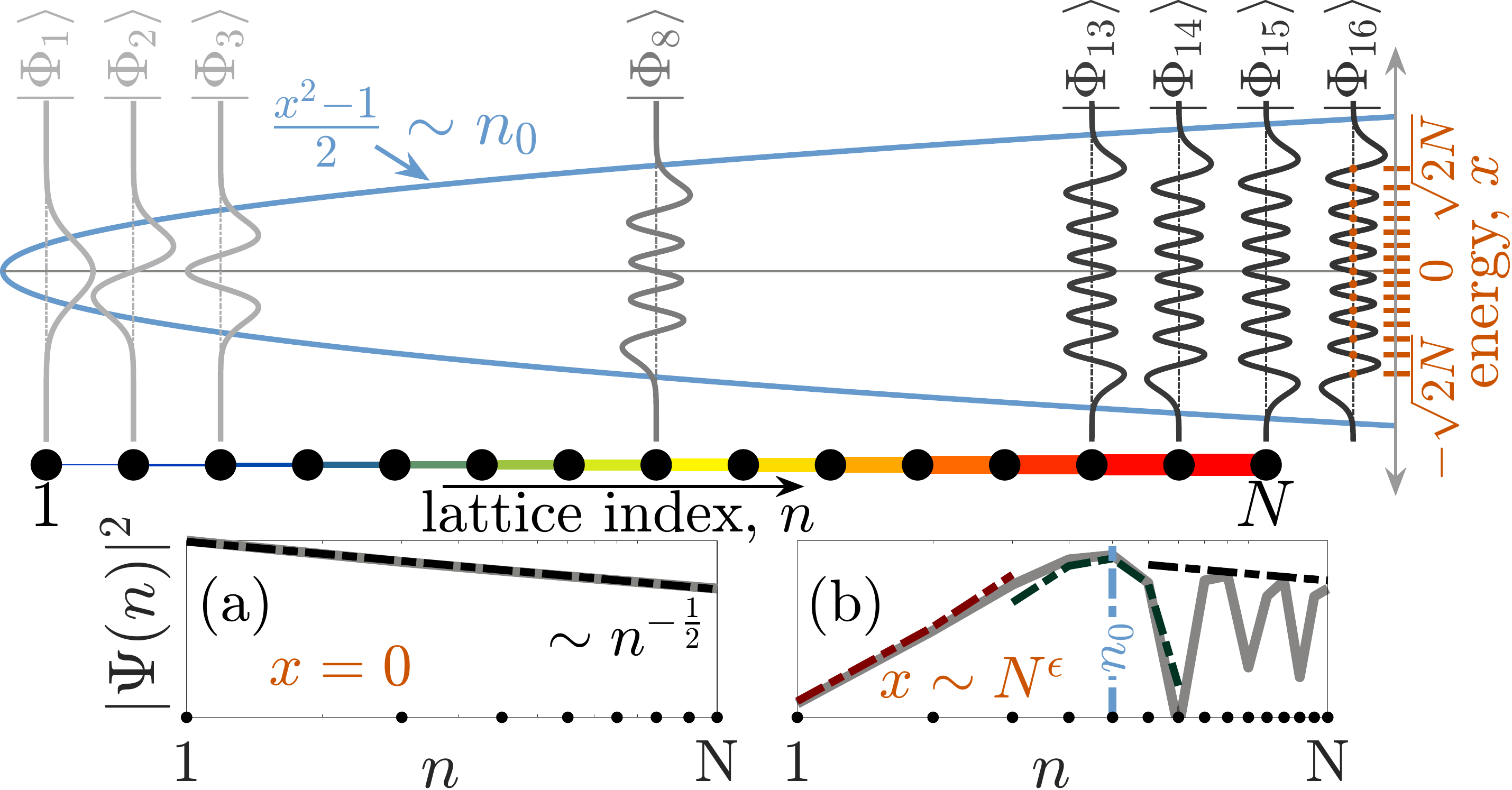}
	\caption{{\bf QHO-to-mean-Hamiltonian correspondence}: QHO eigenstates $\ket{\Phi_n}$ are plotted by gray lines inside the parabola $n = \frac{x^2-1}{2}$. The horizontal QHO energy axis corresponds to the real space $1\leq n\leq N$ of $H_{m,n}$. The eigenenergies of $H_{m,n}$, given by the zeros of $\mathcal{H}_{15}$, lie on the vertical QHO coordinate axis. Colorbar of increasing thickness in the $n$-axis shows the $\sqrt{n}$ growth of the hopping terms.
	Eigenstate intensities of $H_{m,n}$ (a)~at $x=0$, decay as $n^{-\frac{1}{2}}$; (b)~in the spectral bulk $x \sim N^\epsilon$ (shown for $\epsilon = 0.3$), first,
rapidly grow at $n\ll \nmax$, Eq.~\eqref{eq_V_left}, to the left of 
$\nmax$, then have a smooth maximum Eq.~\eqref{eq_V_max_decay1} for $n\approx \nmax$, followed by the power-law decay $\sim n^{-\frac{1}{2}}$ for $n\gg \nmax = \frac{x^2-1}{2}$, Eq.~\eqref{eq_V_right}.
	}
	\label{fig_QHO_cartoon}
\end{figure}

In this Letter, we focus on the case $\beta\to\infty$ reducing to a deterministic tridiagonal {\itshape mean Hamiltonian}
\begin{align}
	\label{eq_H_mean}
	H_{m, n} = \sqrt{\frac{m}{2}} \delta_{m, n-1} + \sqrt{\frac{m - 1}{2}}\delta_{m, n+1}.
\end{align}
The properties of $H_{m, n}$ can be understood by a mapping to the quantum harmonic oscillator (QHO), $\hQHO =\frac{\hat{x}^2 + \hat{p}^2}{2}$ where the $n$th eigenstate $\ket{\Phi_n}$ with energy $(n - \frac{1}{2})$ has the following real-space structure
\begin{align}
	\label{eq_V_QHO}
	 \Phi_n(x) = \pi^{-\frac{1}{4}}e^{-\frac{x^2}{2}} \frac{\mathcal{H}_{n-1}(x)}{\sqrt{2^{n-1} (n-1)!}},\quad n\geq 1,
\end{align}
where $\mathcal{H}_n(x)$ is the $n$th order Hermite polynomial~\cite{GriffithsBook2018}.
The matrix element $H_{m, n} = \bra{\Phi_m} \hat{x} \ket{\Phi_n}$ of the position operator $\hat{x}$ in the eigenbasis ($\{\ket{\Phi_n}\}$) of $\hQHO$ leads to Eq.~\eqref{eq_H_mean},
thus, the energy axis of $H_{m,n}$ maps to the spatial axis $x$ of the QHO and vice versa, as shown in the top panel of Fig.~\ref{fig_QHO_cartoon}.
Hence, we denote the energy of the mean Hamiltonian, $H_{m,n}$ as $E\stackrel{\text{def}}{=} x$ with eigenstates $\ket{\Psi_{x}}$. Imposing the open boundary conditions, $\Psi_x(0) = 0$, $\Psi_x(N+1)=0$, we obtain the recursive relation
\begin{gather}\label{eq_H_V_recur_BC}
	\sqrt{n}\Psi_x(n) + \sqrt{n+1}\Psi_x(n+2) = \sqrt{2}x\Psi_x(n+1) \ .
\end{gather}
Thus, eigenstate components are given by the Hermite polynomials $\mathcal{H}_n(x)$
\begin{align}
	\label{eq_V_mean_H}
	\Psi_{x} (n) = \nrm \frac{\mathcal{H}_{n-1}(x)}{\sqrt{ 2^{n-1} (n-1)! }},\quad n = 1, 2,\dots, N \ ,
\end{align}
where $\nrm \approx e^{-\frac{x^2}{2}} N^{-\frac{1}{4}}$ is the normalization factor~\cite{supple} and the roles of the coordinate $n$ and energy $x$ are interchanged.
The characteristics equation, $\det (H - x\mathbb{I}) = 0$ is resolved by $x_j$, the zeros of the $N$th order Hermite polynomial,
$\mathcal{H}_N(x_j) = 0$, $1\leq j \leq N$.
The largest zero of $\mathcal{H}_N(x)$ is $\sqrt{2N} + \mathcal{O}(N^{-{1}/{6}})$~\cite{SzegoBook1959}, thus the width of the density of states scales as $\mathcal{O}(\sqrt{2N})$ with the system size.

In order to characterize multifractality, similar to Eq.~\eqref{eq:IPR_q}, we introduce the finite-size fractal dimensions $D_q^{(N)}$ via
the $q$th moments of the wave-function intensity $|\Psi_x(n)|^{2}$ in the coordinate basis $n$
\begin{align}
	\label{eq_IPR_q_def}
	\ipr_q^{(N)} = \sum_{n = 1}^{N} |\Psi_x(n)|^{2q} \sim N^{(1-q)D^{(N)}_q}
\end{align}
$\lim\limits_{N\to \infty} D_q^{(N)} \equiv D_q$ is a monotonically decreasing function of $q$~\cite{Soukoulis1984, Hentschel1983, Backer2019} bounded from below by $D_\infty$.
The latter gives the scaling of the maximum eigenstate intensity, $|\Psi_x(\nmax)|^2 \sim N^{-D_\infty}$~\cite{Lakshminarayan2008}, present at the spatial index
\begin{align}
	\label{eq_V_max_loc}
	n\approx \nmax \equiv \frac{x^2-1}{2}\stackrel{\text{def}}{=} N^{2\epsilon}, \quad 0\leq \epsilon \leq \frac12
\end{align}
given by the parabolic potential in QHO, Fig.~\ref{fig_QHO_cartoon}. 
Upon increasing energy $x$ (or $\epsilon$), $\nmax$ grows and approaches the system size $N$ at the spectral edges $x\simeq \pm\sqrt{2N}$.
This leads to the energy-dependent fractal dimensions $D_q$ as we will now show via a systematic study of the system size scaling of the eigenstate intensities.

In order to obtain the eigenstate components near the maximum intensity, we approximate the Hermite polynomials in terms of Airy functions $\Ai{-y}$~\cite{supple}, giving us
\begin{align}
	\label{eq_V_max_decay1}
	\Psi_x(n+1) \approx \frac{\Ai{- \nmax^{-\frac{1}{3}} \eta }}{N^{\frac{1}{4}} \nmax^{\frac{1}{12}} },\quad \eta \equiv n - \nmax\ll \nmax
\end{align}
where $\Ai{-y}$ for an argument $y = \nmax^{-\frac{1}{3}} \eta \sim \mathcal{O}(1)$ has a maximum value $\sim \mathcal{O}(1)$. For $y\gtrsim 1$, $\Ai{-y}$ decays as $\sim y^{-\frac{1}{4}}$ with oscillations as shown in Fig.~\ref{fig_QHO_cartoon}(b). Hence, within $0\leq y\lesssim 1$, equivalently within $1\leq \eta \lesssim n_0^{\frac{1}{3}}\equiv N^{\frac{2\epsilon}{3}}$, we can consider $\Ai{-y}$ to be constant. Such an interval contains $\mathcal{O}\del{N^{\frac{2\epsilon}{3}}}$ eigenstate intensities scaling similar to the maximum intensity
\begin{align}
	\label{eq_V_max_decay2}
	|\Psi_x(\nmax+\eta)|^2 \approx |\Psi_x(\nmax)|^2 \approx N^{-\frac{1}{2} - \frac{\epsilon}{3}}
\end{align}
for a bulk state with energy $x \ll \sqrt{2N}$. Again, $|\Psi_x(\nmax)|^2 \sim N^{-D_\infty}$ along with Eq.~\eqref{eq_V_max_decay2} implies that
\begin{align}
	\label{eq_D_infty_bulk}
	D_\infty = \frac{1}{2} + \frac{\epsilon}{3} \ .
\end{align}
The fractal dimension $D_\infty$ is energy dependent: at the center of the spectrum ($\epsilon=0$), $D_\infty = \frac{1}{2}$ whereas $D_\infty = \frac{2}{3}$ for the spectral bulk, $\epsilon = \frac{1}{2}$.
From Eq.~\eqref{eq_D_infty_bulk}, we get $D_\infty < 1$ implying that the bulk states must be at least non-ergodic, i.e.~they are not uniformly distributed in the $(N-1)$-dim unit hypersphere \cite{Lakshminarayan2008, Vikram2023}.

Next we need to determine whether the non-ergodic bulk states of the mean Hamiltonian are fractal or multifractal. We can explicitly calculate all the fractal dimensions, $D_q$ or infer the multifractality from the spectrum of fractal dimensions (SFD), $f(\alpha)$~\cite{Evers2008, Fyodorov2015, Chhabra1989}. In a multifractal state, there are $\mathcal{O}\del{ N^{f(\alpha)} }$ number of intensities those scale with the system size as $|\Psi|^2\sim N^{-\alpha}$ for a range of $\alpha\geq0$.
The fractal dimension $D_q$ is related to the SFD via the Legendre transform~\cite{Evers2008}
\begin{gather}\label{eq:Legendre}
	(q-1)D_q = \min_\alpha \left[q\alpha - f(\alpha)\right]. 
\end{gather}
Comparing with Eq.~\eqref{eq_D_infty_bulk}, we get $\alpha_{\mathrm{min}} = D_\infty$, while the number of such intensities is $N^{f( \alpha_{\mathrm{min}})}$ with $f( \alpha_{\mathrm{min}} ) = \tfrac{2\epsilon}{3}$.

For $n \ll \nmax$, there is a rapid monotonic growth of the eigenstate components with $n$~\cite{supple}, given by
\begin{align}
	\label{eq_V_left}
	\Psi_x(n+1) &\approx \dfrac{e^{-\frac{x^2}{2}}}{(4\pi N)^{\frac{1}{4}}} x^{-\frac{1}{2}} \del{\dfrac{4x^2}{2n+1}}^{\frac{2n+1}{4}}.
\end{align}
Consequently, there are $\mathcal{O}(1)$ eigenstate components with arbitrarily small intensities for $n \ll \nmax$, leading to $f(\alpha) = 0$ for large values of $\alpha\gg 1$ which do not contribute to $D_q$ for $q>0$~\cite{supple}. We have ignored such $\alpha$ values for our subsequent calculation of the SFD.

On the right side of the maximum, $n\gg \nmax$, we can approximate the eigenstate components at energy $x$ as~\cite{supple}
\begin{align}
	\label{eq_V_right}
	\Psi_x(n+1) \approx \del{\dfrac{2}{\pi N}}^{\frac{1}{4}} \frac{ \cos\del{ \frac{n\pi}{2} - \sqrt{2n+1} x } }{n^{\frac{1}{4}}}.
\end{align}
Hence, eigenstate intensities oscillate with an envelope decaying as $n^{-\frac{1}{2}}$, see Fig.~\ref{fig_QHO_cartoon}(b). In agreement with our principle of observing genuine multifractality in power-law decaying states, the bulk states of the mean Hamiltonian exhibits the power-law decay $n^{-2a}$ with the exponent $2a = \tfrac12 < d$.

\begin{figure}[t]
	\centering
	\includegraphics[width=\columnwidth]{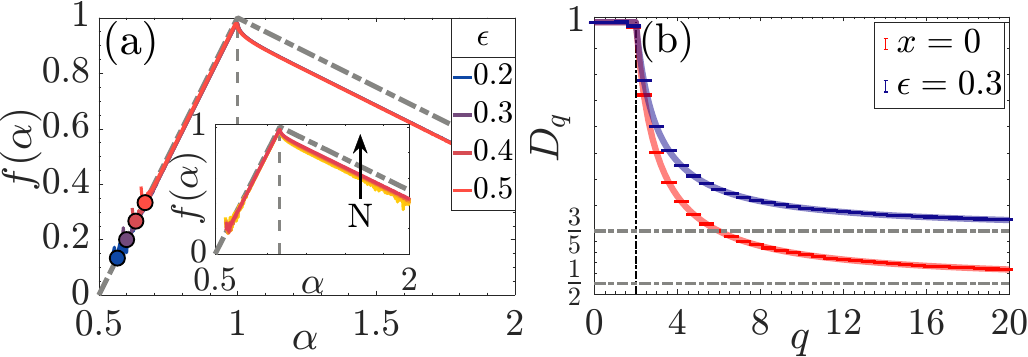}
	\caption{{\bfseries Multifractality of bulk states:} (a)~Spectrum of fractal dimensions for several energies $x = N^{\epsilon}$ for $N = 10^9$. 
		The markers denote $\alpha_{\mathrm{min}} = \frac{1}{2} + \frac{\epsilon}{3}$ and the dashed line denotes the analytical expressions from Eqs.~\eqref{eq_SFD_bulk_1} and \eqref{eq_SFD_bulk_2}.
		The inset shows $f(\alpha)$ for $\epsilon = 0.3$ and different $N$ from $10^6$ to $10^9$. 
		(b)~Fractal dimensions, $D_q$ vs.~$q$ at energies $x = 0$ and $x = N^{0.3}$.
		Solid lines denote analytical expression from Eq.~\eqref{eq_Dq_bulk} and horizontal dashed lines denote $D_\infty$ from Eq.~\eqref{eq_D_infty_bulk}.
	}
	\label{fig_Dq}
\end{figure}

Now we calculate the entire SFD for the bulk eigenstates of the mean Hamiltonian. In Eq.~\eqref{eq_V_right}, the cosine part has a magnitude of order one for most of the values of $n \gg \nmax$. First, we consider the scaling of the envelope where the intensities, $|\Psi_x(n+1)|^2 \sim \del{n N}^{-\frac{1}{2}}\sim N^{-\alpha}$. The number of such components is $n-\nmax \sim n\sim N^{f(\alpha)}$, implying that
\begin{align}
	\alpha = \frac{1}{2} + \frac{f(\alpha)}{2} \Rightarrow f(\alpha) = 2\alpha-1.
\end{align}
It now remains to determine the bounds on the exponent $\alpha$. The upper bound can be obtained by substituting $n\sim N$ in Eq.~\eqref{eq_V_right}. In this limit i.e.~at the tail, $\mathcal{O}(N)$ eigenstate intensities scale as $N^{-1}$, leading to $f(\alpha) = 1$ at $\alpha = 1$. On the other hand, in order to obtain the lower bound, we consider the scaling of the energy as $\nmax \sim N^{2\epsilon}$, $n\gg \nmax$ and $n\sim N^{f(\alpha)}$. It is easy to see that $f(\alpha)>2\epsilon$, leading to the lower bound $\alpha> \frac{1}{2}+\epsilon$. The lower bound can be further refined by considering Eq.~\eqref{eq_V_max_decay1}, which also implies the same power-law decay at $n-\nmax\equiv \eta\gg \nmax^{\frac{1}{3}}$, leading to $f(\alpha) = 2\alpha - 1$ with a left bound on $\alpha$ given by $\eta\sim N^{f(\alpha)}\gg \nmax^{\frac{1}{3}} \Leftrightarrow \alpha>\frac{2\epsilon}{3}$. 
Therefore, the power-law decaying envelope of the bulk states for $n\gtrsim \nmax$ at energy $x\sim N^\epsilon$ leads to the following SFD
\begin{align}
	\label{eq_SFD_bulk_1}
	f(\alpha) = 2\alpha - 1,\quad \frac{1}{2} + \frac{\epsilon}{3} \leq \alpha \leq 1.
\end{align}

Now we consider the small cosine values in Eq.~\eqref{eq_V_right} corresponding to atypically small wave-function components and obtain the full SFD even for $\alpha > 1$. The zeros of the cosine part in Eq.~\eqref{eq_V_right} reside at the lattice indices $\nmax \lesssim n_k \leq N$ such that $\sqrt{2n_k+1}x \approx k\pi$ for odd $n_k$ and $\sqrt{2n_k+1}x \approx k\pi + \frac{\pi}{2}$ for even $n_k$. We focus only on the odd values of $n_k$, as the even case is fully analogous. Since $\nmax \approx x^2 = N^{2\epsilon}$ and $n_k \approx \frac{\pi^2 k^2}{x^2}$, we get
\begin{align}
	\label{eq_V_zeros}
	x^2 \lesssim k \leq \sqrt{N} x \ .
\end{align}
For small deviations $\delta_n = |n - n_k|$, we can approximate the cosine as
\begin{align}
	\del{\sqrt{n_k+\delta_n }-\sqrt{n_k}}x \sim\frac{x \delta_n}{\sqrt{n_k}} \ll 1.
\end{align}
Then, intensities scale as $N^{-\alpha} \sim (N n_k)^{-\frac{1}{2}}\del{\frac{x \delta_n}{\sqrt{n_k}}}^2$ within the interval $[n_k-\delta_n, n_k+\delta_n]$.
As there are $k$ zeros of Eq.~\eqref{eq_V_right}, present up to $n = n_k$, there are $k\cdot \delta_n\sim N^{f(\alpha)}$ components with intensity $\sim N^{-\alpha}$.
Substituting $\delta_n$ from one expression to another and taking into account that $\sqrt{n_k}x\sim k$, one finds that
\begin{align}
	\label{eq:SFD_small_cos}
	f(\alpha) = \frac{1}{4} - \frac{\alpha}{2} + \frac{5}{2} \frac{\log \frac{k}{x}}{\log N}.
\end{align}
Eq.~\eqref{eq:SFD_small_cos} is dominated by the maximum value of $k = \sqrt{N} x$, Eq.~\eqref{eq_V_zeros}. Therefore, the atypically small intensities of the bulk states for $n\gtrsim \nmax$ at energy $x\sim N^\epsilon$ produces the SFD
\begin{align}
	\label{eq_SFD_bulk_2}
	f(\alpha) = \dfrac{3-\alpha}{2},\quad 1 \leq \alpha.
\end{align}
In Fig.~\ref{fig_Dq}(a), we show the numerical estimate of the entire SFD for different $N$ and $\epsilon$, giving an excellent agreement with Eqs.~\eqref{eq_SFD_bulk_1} and \eqref{eq_SFD_bulk_2}.

The Legendre transform, Eq.~\eqref{eq:Legendre}, of $f(\alpha)$ from Eqs.~\eqref{eq_SFD_bulk_1} and~\eqref{eq_SFD_bulk_2} gives the fractal dimensions
\begin{align}
	\label{eq_Dq_bulk}
	\begin{split}
		D_q &= \begin{cases}
			1, & 0 \leq q \leq 2\\
			1-\left(\dfrac12 - \dfrac{\epsilon}{3}\right)\dfrac{q-2}{q-1}, & \phantom{0 \leq }\;q > 2.
		\end{cases}
	\end{split}
\end{align}
which is in the full agreement with our generic result Eq.~\eqref{eq:IPR_q_multifractal} for $a=\frac{1}{4}$ and $d=1$. The limit $q\to\infty$ in the above expression retrieves $D_\infty$ given in Eq.~\eqref{eq_D_infty_bulk}.
In Fig.~\ref{fig_Dq}(b), the numerically estimated $D_q$ vs.~$q$ for eigenstates at energies $x \sim N^{0.3}$ and $x = 0$ show excellent agreement with the analytical expression from Eq.~\eqref{eq_Dq_bulk}. Importantly we show the $q$-dependence of $D_q$ implying multifractality in the eigenstate fluctuations.

To summarize, in this Letter we provide a simple, but powerful principle to realize genuine multifractality as a quantum phase. The necessary criteria for such a phase is to have a sufficiently slow power-law decay of the eigenstates with an exponent at least two times smaller than the lattice dimensionality.
We discuss why in most of the long-range models such wave-function power-law decay cannot be realized and only fractal phases of matter can be observed. As an illustration of our principle, we provide a short-range 1d Hamiltonian where genuine multifractal states with the power-law decaying profile are realized and analytically find their fractal dimensions.
As the considered model is the $\beta\to\infty$ limit of the \bte, we expect that predicted multifractality will survive for finite $\beta>1$ at least until the eigenstates  localize~\cite{Breuer2007,breuer2010spectral}, but we keep the detailed investigation of these cases for our future studies.

A.~K.~D. is supported by an INSPIRE Fellowship, DST, India and the Fulbright-Nehru grant no.~2879/FNDR/2023-2024. I.~M.~K. acknowledges the support 
of the European Research Council under the European Union's Seventh Framework Program Synergy ERC-2018-SyG HERO-810451.

\clearpage 
\appendix
\renewcommand\thefigure{\thesection.\Roman{figure}}
\setcounter{figure}{0}
\renewcommand\thetable{\thesection.\Roman{table}}
\setcounter{table}{0}
\renewcommand\theequation{\Roman{equation}}
\setcounter{equation}{0}
\onecolumngrid
\section{Asymptotes of the Hermite polynomials}
Following the approximation of Hermite polynomials~\cite{SzegoBook1959, Dominici2006}, the eigenstate components of the mean Hamiltonian at energy $x$ can be expressed as
\begin{align}
	\label{eq_Apnd_V_apprx}
	\begin{split}
		\Psi_x(n+1) &\approx \begin{cases}
			C_x n^{-\frac{1}{4}} \del{\dfrac{x^2}{2n+1} - 1}^{-\frac{1}{4}} \exp\del{ \del{n+\dfrac{1}{2}} \del{ \cosh^{-1} \dfrac{x}{\sqrt{2n+1}} - \dfrac{x}{\sqrt{2n+1}} \sqrt{\dfrac{x^2}{2n+1} - 1} } }, & n < \nmax\\
			\dfrac{\nrm (2\pi)^{\frac{1}{4}}}{n^{\frac{1}{12}}} \exp\del{\sqrt{2n} x - n} \Ai{\sqrt{2} (x - \sqrt{2n}) n^{\frac{1}{6}}}, & n\approx \nmax\\
			C_x n^{-\frac{1}{4}} \del{1 - \dfrac{x^2}{2n+1}}^{-\frac{1}{4}} \sin \del{ \dfrac{3\pi}{4} + \del{n+\dfrac{1}{2}}\del{ \dfrac{x}{\sqrt{2n+1}}\sqrt{1-\dfrac{x^2}{2n+1}} - \cos^{-1}\dfrac{x}{\sqrt{2n+1}} } }, & n > \nmax
		\end{cases}
	\end{split}
\end{align}
where $\nmax \equiv \frac{x^2-1}{2}$, $C_x = \frac{\nrm e^{\frac{x^2}{2}}}{(8\pi)^{\frac{1}{4}}}$ if $n<\nmax$ and $C_x = \frac{\nrm e^{\frac{x^2}{2}}}{(\pi/2)^{\frac{1}{4}}}$ otherwise and $\nrm$ is the eigenstate normalization factor.

\section{Eigenstate components for $n \ll \nmax$}
Eq.~\eqref{eq_Apnd_V_apprx} implies that for $n\ll \nmax$,
\begin{align}
	\label{eq_Apnd_V_apprx_left}
	\begin{split}
		\Psi_x(n+1) &\approx C_x n^{-\frac{1}{4}} \frac{(2n+1)^{\frac{1}{4}}}{\sqrt{x}} \exp\del{ \del{n+\dfrac{1}{2}} \del{ \cosh^{-1} \dfrac{x}{\sqrt{2n+1}} - \dfrac{x}{\sqrt{2n+1}} \sqrt{\dfrac{x^2}{2n+1} - 1} } } \\
		&\approx \frac{C_x}{\sqrt{x}} n^{-\frac{1}{4}} (2n+1)^{\frac{1}{4}} \exp\del{ \del{n+\dfrac{1}{2}} \del{ \ln \frac{2x}{\sqrt{2n+1}} - \frac{x^2}{2n+1} } }\\
		&= \frac{C_x}{\sqrt{x}} n^{-\frac{1}{4}} (2n+1)^{\frac{1}{4}} \del{e^{-\frac{x^2}{2n+1}} \frac{2x}{\sqrt{2n+1}} }^{n+\frac{1}{2}} = \frac{C_x}{\sqrt{x}} e^{-\frac{x^2}{2}}\frac{(2x)^{n+\frac{1}{2}}}{ n^{\frac{1}{4}} (2n+1)^{\frac{n}{2}} } = \dfrac{\nrm e^{\frac{x^2}{2}}}{ (8\pi)^{\frac{1}{4}} \sqrt{x}} e^{-\frac{x^2}{2}}\frac{(2x)^{n+\frac{1}{2}}}{ n^{\frac{1}{4}} (2n+1)^{\frac{n}{2}} }\\
		\Rightarrow\Psi_x(n+1) &\approx \frac{\nrm}{(4\pi)^{\frac{1}{4}}} x^{-\frac{1}{2}} \del{\frac{4x^2}{2n+1}}^{\frac{2n+1}{4}}
	\end{split}
\end{align}
Since $\dfrac{x^2}{2n+1}\gg 1$, $\Psi_x(n+1)$ is a monotonically increasing function of $n$ for $n\ll \nmax$.

\section{Eigenstate components for $n \gg \nmax$}
Let, $m = \dfrac{x}{\sqrt{2n+1}}$. Then, for $n\gg \nmax$ (where $C_x = \frac{\nrm e^{\frac{x^2}{2}}}{(\pi/2)^{\frac{1}{4}}}$)
\begin{align}
	\label{eq_Apnd_V_apprx_right}
	\begin{split}
		\Psi_x(n+1) &\approx C_x n^{-\frac{1}{4}} \del{1 - m^2}^{-\frac{1}{4}} \cos\del{ \frac{\pi}{4} + \del{n+\dfrac{1}{2}} \del{ m \del{1 - m^2}^{\frac{1}{2}} - \cos^{-1}m }  }\\
		&\approx C_x n^{-\frac{1}{4}} \del{1 + \frac{m^2}{4}} \cos\del{ \frac{\pi}{4} + \del{n+\dfrac{1}{2}} \del{ m - \frac{m^3}{2} - \frac{\pi}{2} + m + \frac{m^3}{6} }  }\\
		&\approx C_x n^{-\frac{1}{4}} \cos\del{ -\frac{n\pi}{2} + m(2n+1) }
		\approx \begin{cases}
			(-1)^m C_x n^{-\frac{1}{4}} \cos\del{\sqrt{2n + 1} x}, & n = 2m\\
			(-1)^m C_x n^{-\frac{1}{4}} \sin\del{\sqrt{2n + 1} x}, & n = 2m+1
		\end{cases}
	\end{split}
\end{align}
Therefore, the eigenstate components oscillate and the envelope decays as $n^{-\frac{1}{4}}$ for $n\gg \nmax$.

\section{Normalization constant of the eigenstate}
As the eigenstate components monotonically increase for $n \ll \nmax$ and decay as $n^{-\frac{1}{4}}$ for $n \gg \nmax$, the maximum intensity occurs at $n = \nmax \approx \dfrac{x^2-1}{2}$. If we take a bulk energy $x\ll \sqrt{2N}$, then
\begin{align}
	\label{eq_Apnd_V_bulk_nrm}
	\begin{split}
		\sum_{n = 0}^{N-1} |\Psi_x(n+1)|^2 &\approx \frac{\nrm ^2 e^{x^2}}{\sqrt{\pi/2}} \sum_{n = \nmax+1}^{N-1} n^{-\frac{1}{2}} \approx \frac{\nrm ^2 e^{x^2}}{\sqrt{\pi/2}} \del{ h\del{N-1, \frac{1}{2}} - h\del{\nmax, \frac{1}{2}} } \approx \frac{\nrm ^2 e^{x^2}}{\sqrt{\pi/2}} 2 \del{ \sqrt{N} - \sqrt{\nmax} }\\
		\Rightarrow \nrm &\approx e^{-\frac{x^2}{2}} N^{-\frac{1}{4}}
	\end{split}
\end{align}
where $\zeta(x)$ is the Riemann zeta function and $h(n, r)$ = $n$th harmonic number of order $r$ approximated using Euler-Maclaurin sum formula ($h(n, \frac{1}{2}) = \zeta\del{\frac{1}{2}} + 2\sqrt{n} + \mathcal{O}(n^{-\frac{1}{2}})$).

\section{Eigenstate components around the maximum intensity}
Following the approximation of the Hermite polynomials for $x\to \sqrt{2n}$ and $n\to \infty$~\cite{Dominici2006}, we get
\begin{align}
	\label{eq_Apnd_V_apprx_nmax}
	\begin{split}
		\Psi_x(n+1) &= (2\pi)^{\frac{1}{4}} N^{-\frac{1}{4}} n^{-\frac{1}{12}} \exp\del{ \sqrt{2n}x - n - \frac{x^2}{2}} \Ai{ \sqrt{2}(x - \sqrt{2n}) n^{\frac{1}{6}} }
	\end{split}
\end{align}
where $\Ai{y}$ is the Airy function. Let, $n = \nmax + \eta,\: \eta\ll \nmax$ such that the approximation in Eq.~\eqref{eq_Apnd_V_apprx_nmax} is valid for $\nmax \lessapprox n\lessapprox \nmax + \eta$. Then, we can simplify Eq.~\eqref{eq_Apnd_V_apprx_nmax} as
\begin{align}
	\label{eq_Apnd_V_apprx_middle}
	\begin{split}
		\Psi_x(n+1) &\approx N^{-\frac{1}{4}} (\nmax + \eta)^{-\frac{1}{12}} \exp\del{ \sqrt{2(\nmax + \eta)}x - (\nmax + \eta) -\frac{x^2}{2} } \Ai{ \sqrt{2}(x - \sqrt{2(\nmax + \eta)}) (\nmax + \eta)^{\frac{1}{6}} }\\
		&\approx N^{-\frac{1}{4}} \nmax^{-\frac{1}{12}} \exp\del{ 2\nmax\del{1 + \frac{\eta}{2\nmax}} - 2\nmax - \eta } \Ai{ 2 \sqrt{\nmax} \del{1 - \sqrt{1 + \frac{\eta}{\nmax}}} \nmax^{\frac{1}{6}} \del{1 + \frac{\eta}{\nmax}}^{\frac{1}{6}} }\\
		&\approx N^{-\frac{1}{4}} \nmax^{-\frac{1}{12}} \Ai{ - \nmax^{-\frac{1}{3}} \eta }
	\end{split}
\end{align}

Combining Eqs.~\eqref{eq_Apnd_V_apprx_left}, \eqref{eq_Apnd_V_apprx_right} and \eqref{eq_Apnd_V_apprx_middle} we get
\begin{align}
	\label{eq_Apnd_V_decay}
	\Psi_{x}(n+1) \approx \begin{cases}
		\dfrac{e^{-\frac{x^2}{2}}}{(4\pi N)^{\frac{1}{4}}} x^{-\frac{1}{2}} \del{\dfrac{4x^2}{2n+1}}^{\frac{2n+1}{4}}, & n\ll \nmax\\
		N^{-\frac{1}{4}} \nmax^{-\frac{1}{12}} \Ai{ - \nmax^{-\frac{1}{3}} \eta }, & n\approx \nmax\\
		\del{\dfrac{2}{\pi N}}^{\frac{1}{4}} n^{-\frac{1}{4}} \cos\del{ \dfrac{n\pi}{2} - \sqrt{2n+1} x }, & n\gg \nmax
	\end{cases}
\end{align}

\section{Spectrum of fractal dimensions for $n \gg \nmax$}

Eq.~\eqref{eq_Apnd_V_apprx_left} implies that for $n\ll \nmax$ 
\begin{align}
	\begin{split}
		\Psi_x(n+1) &\approx \dfrac{ N^{-\frac{1}{4}} e^{-\frac{x^2}{2}} }{(4\pi)^{\frac{1}{4}}} x^{-\frac{1}{2}} \del{\dfrac{4x^2}{2n+1}}^{\frac{2n+1}{4}}\\
		\Rightarrow N^{-\alpha} &= N^{-\frac{1}{2}} e^{-x^2} x^{-1} \del{\frac{4x^2}{2n+1}}^{\frac{2n+1}{2}} = N^{-\frac{1}{2}} e^{-N^{2\epsilon}} N^{-\epsilon} \del{\frac{4N^{2\epsilon}}{2n+1}}^{\frac{2n+1}{2}}\\
		\Rightarrow N^{\epsilon + \frac{1}{2} - \alpha} &= e^{-N^{2\epsilon}} \del{\frac{4N^{2\epsilon}}{2n+1}}^{\frac{2n+1}{2}}\Rightarrow \del{\epsilon + \frac{1}{2} - \alpha} \ln N \approx -N^{2\epsilon} - n\ln\del{\frac{n}{N^{2\epsilon}}}\\
		\Rightarrow \frac{n}{N^{2\epsilon}} \ln\del{\frac{n}{N^{2\epsilon}}} &= \del{\alpha - \epsilon -\frac{1}{2}} \frac{\ln N}{N^{2\epsilon}} - 1
	\end{split}
\end{align}
Let, $y = \dfrac{n}{N^{2\epsilon}}$. For $\alpha > \epsilon + \dfrac{1}{2}$, $y\log y\approx \mathcal{O}\del{\dfrac{\ln N}{N^{2\epsilon}}}$ and $0 < y < 1$ as $1\leq n<N^{2\epsilon}$, where $-e^{-1} \leq y\ln y < 0$ ($y\ln y = -e^{-1}$ for $y = y_0\approx 0.36788$). Consider the domain $[n, n+\delta_n]$ within which $y - y_0 \approx y_0\ln y_0$, i.e.
\begin{align}
	y - y_0 &= \frac{n+\delta_n}{N^{2\epsilon}} - \frac{n}{N^{2\epsilon}} = \frac{\delta_n}{N^{2\epsilon}} \approx \dfrac{\ln N}{N^{2\epsilon}} \Rightarrow \delta_n \approx \ln N
\end{align}

Because $y\ln y<0$, we also require that
\begin{align}
	\dfrac{\ln N}{N^{2\epsilon}} < 1\Rightarrow \dfrac{\ln \ln N}{\ln N} < 2\epsilon, \qquad \lim\limits_{N\to \infty} \dfrac{\ln \ln N}{\ln N} = 0
\end{align}
However, $\dfrac{\ln \ln N}{\ln N}$ converges very slowly to 0, hence for finite $N$, we cannot consider $\epsilon$ very close to 0. Then,
\begin{align}
	\del{\epsilon + \frac{1}{2}} + \frac{(1-e^{-1}) N^{2\epsilon}}{\ln N} \leq \alpha \leq \del{\epsilon + \frac{1}{2}} + \frac{N^{2\epsilon}}{\ln N}
\end{align}
Therefore, left side of $\nmax$ contributes arbitrarily small intensities, i.e.~$N^{-\alpha}$ with $\alpha\gg 1$, and corresponding singularity spectrum is
\begin{align}
	f(\alpha) = \frac{\ln \ln N}{\ln N} \sim 0
\end{align}
Such small intensities are well below the machine precision for a modest value of $N\sim 10^6$.

\bibliography{ref_multifractal}
\end{document}